\documentclass[aps,prl,twocolumn,showpacs,amsmath,amssymb,superscriptaddress]{revtex4-1}

\usepackage{graphicx}

\begin{document}

\title{Multicriticality, Metastability, and Roton Feature in Bose-Einstein Condensates with
Three-Dimensional Spin-Orbit Coupling}

\author{Renyuan Liao}
\affiliation{College of Physics and Energy, Fujian Normal University, Fuzhou 350117, China}

\author{Oleksandr Fialko}
\affiliation{Institute of Natural and Mathematical Sciences and Centre for Theoretical Chemistry
and Physics, Massey University Auckland, Private Bag 102904, North Shore, Auckland 0745,
New Zealand}

\author{Joachim Brand}
\affiliation{New Zealand Institute for
Advanced Study, Dodd-Walls Centre for Photonics and Quantum Technology, and Centre for Theoretical Chemistry and Physics, Massey University Auckland,
Private Bag 102904, North Shore, Auckland 0745, New Zealand}

\author{Ulrich Z\"ulicke}
\affiliation{School of Chemical and Physical Sciences and Dodd-Walls Centre for Photonics and
Quantum Technology, Victoria University of Wellington, PO Box 600, Wellington 6140, New Zealand}

\date{\today}

\begin{abstract}
We theoretically study homogeneously trapped atomic Bose-Einstein condensates where all three momentum components couple to a
pseudo-spin-$1/2$ degree of freedom.
Tuning the anisotropies of spin-orbit coupling and the spin-dependent interactions is shown to
provide access to a rich phase diagram with a tetracritical point, first-order phase transitions,
and multiple metastable phases of stripe and plane-wave character. The elementary excitation spectrum of the axial plane-wave phase features an anisotropic roton feature and can be used to probe the phase diagram. In
addition to providing a versatile laboratory for studying fundamental concepts in statistical physics, the
emergence of metastable phases creates new opportunities for observing false-vacuum decay and
bubble nucleation in ultra-cold-atom experiments.
\end{abstract}

\pacs{67.85.Fg, 03.75.Mn, 05.30.Jp, 67.85.Jk}

\maketitle

The possibility to create artificial gauge fields in neutral ultra-cold atom systems~\cite{DAL11,GOL14}
has drastically expanded the array of possibilities for highly controlled experimental simulation of quantum
many-particle systems~\cite{BLO12,GOL14}. In particular, it has become possible to explore effects
associated with spin-orbit coupling (SOC)~\cite{VIC13} that give rise to intriguing phenomena such as
the quantum spin Hall effect~\cite{KAN05,BER06a,BER06b,KON07}, new materials classes such as
topological insulators and superconductors~\cite{BER06b,KON07,HAS10,QI11}, and exotic quasiparticle
excitations such as Majorana fermions~\cite{ALI12,BEE13,ELL15}. In bosonic-atom systems, the
presence of SOC was found to generate novel ground states that have no known analogues in
conventional solid-state materials~\cite{STA08,ZHA10,TIN11}. Intense theoretical attention has focused
on the many-body physics of spin-orbit-coupled Bose-atom systems in free
space~\cite{BAR12,SED12,BAY12,YUN12,CUI13,LIA13,FIA14} and in harmonic
traps~\cite{SIN11,XU12,HU12,KAW12,UED12,GOU12,CON13}.
To date, only a special type of SOC involving a single Cartesian component of the atoms' momentum
has been realized in the lab~\cite{LIN11,CHE12,ZHA12,ZWI12}. However, several proposals exist for
creating a Rashba-type SOC in higher dimensions~\cite{AND12,LI12,XU13}. Experimental
progress is spurred by predictions of an exotic half-vortex phase~\cite{SIN11,HU12} and a striped
superfluid phase~\cite{ZHA10} is systems with two-dimensional Rashba SOC. The three-dimensional
analogue of Rashba SOC is also interesting because it is expected to stabilize a long-sought
Skyrmion mode in the ground state of trapped Bose-Einstein condensates~\cite{KAW12,CON13}. The
intriguing possibility to simulate the so-called Weyl SOC has also been suggested~\cite{AND12}.

Despite the rapid pace of theoretical and experimental studies in the field of spin-orbit-coupled atom
gases, the physical properties of an extended interacting Bose system in the presence of
three-dimensional (3D) SOC have not yet been considered in any detail. This clearly presents a crucial
gap in our basic understanding, as the extended system's behavior  constitutes an important benchmark for
identifying effects associated solely with trapping potentials. Furthermore, real experimental systems
can be designed  with a flat-bottom potential to approximate the extended system and give access to the intriguing
physics demonstrated by our present study \cite{Gaunt2013}. In particular, we show that 3D SOC in an interacting Bose
gas leads to a highly nontrivial phase diagram featuring a tetracritical point, first-order phase transitions,
and emergent metastable phases -- none of which have been seen in systems with lower-dimensional
SOC and/or in the presence of a trapping potential. Thus this system provides opportunities to study
ramifications of multicriticality~\cite{AHA03} and metastability, including false-vacuum decay and bubble
nucleation \cite{Langer1967a,COL77,Schmelzer2006,Fialko2014}, in ultra-cold-atom experiments. We have also studied the
spectrum of elementary excitations and find it to be useful for probing the multitude of
phases and phase transitions.

\textit{The model.\/} ---
We consider a 3D homogeneous interacting two-component Bose gas  subject to cylindrically symmetric
spin-orbit coupling, described by the Hamiltonian $H = H_0 + H_\mathrm{I}$, with
\begin{subequations}
\begin{eqnarray}
H_0 &=& \int \! d^3 r \,\, \Psi^\dagger(\mathbf{r}) \left[ \frac{\mathbf{\hat p}^2}{2m} + \lambda \left(
\hat{\sigma}_\perp \cdot \mathbf{\hat{p}}_\perp + \gamma\, \hat{\sigma}_z \, \hat{p}_z \right)\right]
\Psi(\mathbf{r}) \,\, , \nonumber \\ \\
H_\mathrm{I} &=& \int \! d^3 r \,\, \left[ g\sum_\sigma n_\sigma^2(\mathbf{r}) + 2 g_{\uparrow
\downarrow}\, n_\uparrow(\mathbf{r}) \, n_{\downarrow}(\mathbf{r}) \right] \quad .
\end{eqnarray}
\end{subequations}
Here $\Psi(\mathbf{r})=(\psi_\uparrow,\psi_\downarrow)^T$ is a two-component spinor field,
$n_\sigma=\psi^\dagger_\sigma\psi_\sigma$ is the density for component $\sigma\in \{ \uparrow,
\downarrow\}$, $m$ is the atomic mass, $\hat{\sigma}_j$ (with $j=x,y,z$) denote the Pauli matrices
and $\hat{p}_j=-i\hbar\hat{\nabla}_j$ are the Cartesian components of the single-atom momentum
operator $\mathbf{\hat p}$.

The parameter $\lambda$ measures the SOC strength involving the momentum
$\mathbf{\hat p}_\perp$ in the $xy$ plane, and the dimensionless number $\gamma$
describes the anisotropy of SOC for the momentum component parallel to the $z$ direction.
Note that the limit $\gamma=0$ is the unitary equivalent of the conventional Rashba SOC~\cite{DAL11},
$\gamma=1$ realises the so-called Weyl SOC~\cite{AND12}, and a situation corresponding to the
experimentally created Rashba-type SOC~\cite{LIN11,CHE12,ZWI12,ZHA12} is obtained when
$\gamma\to\infty$ (with $\lambda \gamma$ finite). Fundamentally, the parameter $\gamma$ could be tuned by a sequence of pulsed
inhomogeneous magnetic fields~\cite{AND12}. Also, the strength $g$ ($g_{\uparrow\downarrow}$) of
interactions between same-spin (opposite-spin) components can be varied using an appropriate
Feshbach resonance~\cite{CHE10}. In the special case when $\gamma=1$ and $g=g_{\uparrow
\downarrow}$, the Hamiltonian $H$ is symmetric with respect to simultaneous rotations
of the internal pseudo-spin-$1/2$ degree of freedom and the particle momentum. Throughout the
rest of the paper, we use units such that $\hbar=k_\mathrm{B}=2m=1$.

Diagonalization of $H_0$ yields the two-branch single-particle energy spectrum $E_{\pm}(\mathbf{p})
=(p\pm\lambda\tau_{\mathbf p}/2)^2-\lambda^2\tau_{\mathbf{p}}^2/4$ as a function of 3D momentum
$\mathbf{p}$. Using spherical coordinates, $\mathbf{p}\equiv (p\sin
\theta_\mathbf{p}\cos\varphi_\mathbf{p}, p\sin\theta_\mathbf{p}\sin\varphi_\mathbf{p}, p\cos
\theta_\mathbf{p})$, we have $\tau_{\bf{p}}=\sqrt{\gamma^2 \cos^2{\theta_\mathbf{p}}+
\sin^2{\theta_\mathbf{p}}}$, and the eigenspinors are given by
\begin{equation}
\Psi_\pm(\mathbf{p})=\left( \begin{array}{c} \frac{\pm \sin{\theta_\mathbf{p}}}{\sqrt{\tau_\mathbf{p}
\mp \gamma\cos\theta_\mathbf{p}}} \,\, e^{-i \varphi_\mathbf{p}} \\[0.3cm] \sqrt{\tau_{\bf p} \mp
\gamma \cos{\theta_\mathbf{p}}} \end{array}\right) \,\, \frac{e^{i\mathbf{p}\cdot\mathbf{r}}}{\sqrt{2
\tau_\mathbf{p}}} \quad .
\end{equation}
The lowest-energy state for a given propagation direction parameterized by $\theta_\mathbf{q}$
and $\varphi_\mathbf{q}$ is from the ``$-$" branch and occurs at the momentum $\mathbf{q}$ satisfying
$q = \lambda\tau_{\bf q}/2$.

\textit{The phase diagram.\/} ---
To determine the ground state of the interacting system, as it is routinely done in the
literature~\cite{ZHA10,YUN12}, we assume that the system has condensed into a coherent
superposition of two plane-wave states with momenta $\pm{\mathbf q}$ having magnitude
$q=\lambda\tau_{\mathbf q}/2$. Thus the condensate wave function has the form $\Phi_0
= C_+\, \Psi_- ({\mathbf q}) + C_-\, \Psi_-(-{\mathbf q})$, with coefficients $C_\pm$ that will
be determined by a variational procedure. The condition $n_0=|C_+|^2+|C_-|^2$, with $n_0$
being the particle number density, suggests the parameterization $|C_-|^2=n_0\cos^2{(\alpha/2)}$
and $|C_-|^2=n_0\sin^2{(\alpha/2)}$, with $\alpha\in[0,\pi]$. Introducing the dimensionless
nonlinear-coupling parameter $\tilde{g}\equiv(g_{\uparrow\downarrow}-g)n_0/\lambda^2$,
we find the variational ground-state energy density $E_\mathrm{g}\equiv\langle \Phi_0|H|\Phi_0
\rangle/\mathcal{V}-gn_0$ given by
\begin{equation}\label{Eq.Eg}
\frac{E_\mathrm{g}}{\lambda^2n_0}=-\frac{\tau_{\bf q}^2}{4}+\frac{\tilde{g}}{2}\left[\sin^2{\alpha}
\left(1-\frac{3\sin^2{\theta_\mathbf{q}}}{2\tau_{\bf q}^2}\right) +\frac{\sin^2{\theta_\mathbf{q}}}
{\tau_{\bf q}^2}\right] .
\end{equation}

We first consider the familiar case of two-dimensional SOC by setting $\gamma=0$. Minimization of
$E_\mathrm{g}$ with respect to $\theta_{\bf q}$ and $\alpha$ then yields $\sin^2{\theta_\mathbf{q}}=1$
and $\sin^2{\alpha}=1 \, (0)$ for $\tilde{g}>0$ ($\tilde{g}<0$). The first condition implies that the
condensation momentum is pinned in the $xy$ plane, and the latter condition yields the stripe
phase for $\tilde{g}>0$ ($|C_+|=|C_-|$, i.e. condensation in a coherent superposition of the
opposite-momentum states) or the plane wave phase for $\tilde{g}<0$ (either $|C_+|=0$ or $|C_-|=0$,
i.e. condensation at only one momentum eigenstate). We thus reproduce the ground-state structure of
the conventional two-dimensional-Rashba SOC case~\cite{ZHA10}.

\begin{figure}[t]
\includegraphics[width=0.9\columnwidth]{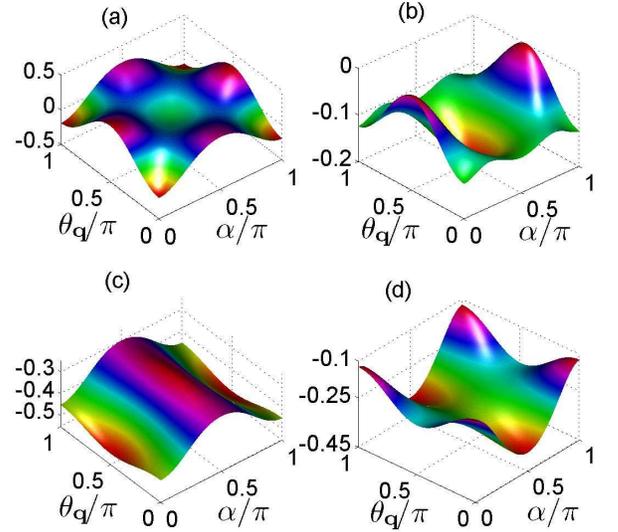}
\caption{(color online) The ground-state energy density $E_\mathrm{g}$ (measured in terms of $\lambda^2n_0$) plotted as a function of the
variational parameters $\theta_\mathbf{q}$ and $\alpha$, for particular values of quantities
characterising anisotropy of nonlinear interactions ($\tilde g$) and spin-orbit coupling ($\gamma$): (a)
$\tilde{g}=1.0$ and $\gamma^2=0.8$; (b) $\tilde{g}=0.25$ and $\gamma^2=0.5$; (c) $\tilde{g}=-0.1$
and $\gamma^2=1.8$; and (d) $\tilde{g}=-0.25$ and $\gamma^2=0.5$. The global minimum in each
panel corresponds to a true ground state, while the existence of local minima on the energy landscape
signifies the emergence of metastable phases.}
\label{fig1}
\end{figure}

Setting $\gamma\ne 0$ unpins the condensation momentum from the $xy$ plane, making it possible
to condense into a state whose momentum has a finite $z$ component. In the following, we will term
such condensation as ``Polar", while condensation into a momentum that lies in the $xy$ plane will be
called ``Axial". As each of these cases can support a stripe (SP) or plane-wave (PW) condensate,
depending on the interaction strength $\tilde{g}$, we have four distinct possible phases: PW-Axial,
SP-Axial, PW-Polar, and SP-Polar. Examination of the variational ground-state energy landscape
shows that each of the four phases is found to be either a true ground state or a metastable state,
depending on the values of $\tilde{g}$ and $\gamma$. See Fig.~\ref{fig1} for pertinent examples.
The stability of the phases can be ensured by the positivity of the Hessian matrix
\begin{equation}
   h_E=\begin{pmatrix}
       \frac{\partial^2 E_g}{\partial \theta_\mathbf{q}^2} &
       \frac{\partial E_g}{\partial\theta_\mathbf{q}}\frac{\partial E_g}{\partial\alpha}\\
       \frac{\partial E_g}{\partial\theta_\mathbf{q}}\frac{\partial E_g}{\partial\alpha}&
       \frac{\partial^2 E_g}{\partial \alpha^2}
   \end{pmatrix} \quad .
\label{Eq.hessian}
\end{equation}

\begin{figure}[t]
\includegraphics[width=0.75\columnwidth]{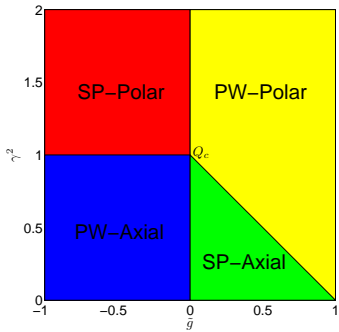}
\caption{(color online) Phase diagram controlled by varying the parameter $\tilde{g}$ that
measures anisotropy of spin-dependent interaction strengths and the quantity $\gamma^2$ related
to anisotropy of spin-orbit coupling. Four possible phases exist -- PW-Polar, SP-Polar, PW-Axial, and
SP-Axial -- that intersect at the tetracritical point $Q_\mathbf{c}$. For Polar (Axial) phases, the
condensate momentum has a finite (vanishing) $z$ component. In the PW (SP) phases,
condensation occurs into a single (a superposition of two) plane-wave state(s).}
\label{fig2}
\end{figure}

The true ground state phase diagram spanned by tuning parameters $\tilde{g}$ and $\gamma^2$ is
shown in Fig.~\ref{fig2}. A tetracritical point $Q_\mathrm{c}$ connecting the four possible phases
emerges when $\tilde{g}=0$ and $\gamma=1$. At this high-symmetry point, the system is invariant with
respect to simultaneous SU(2) spin rotation and rotation of the momentum of the atoms. The observed
behavior at $Q_\mathrm{c}$ in our system contrasts with that exhibited in the presence of a tight harmonic
trapping potential where Skyrmion textures are stabilized in the ground state~\cite{KAW12}.

\begin{figure}[t]
\includegraphics[width=0.75\columnwidth]{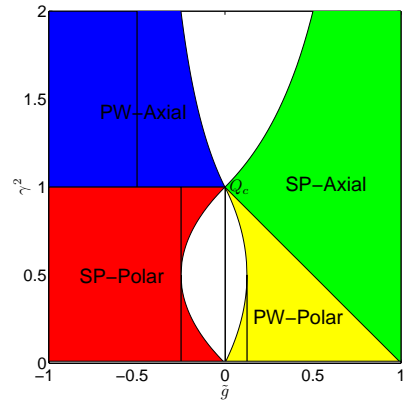}
\caption{(color online) The phase diagram of metastable states controlled by $\tilde{g}$ and
$\gamma^2$. White regions indicate parameter combinations for which there no dynamically
stable metastable phases exist, i.e., where the Hessian matrix in Eq.~(\ref{Eq.hessian}) is not
positive-definite.}
\label{fig3}
\end{figure}

The lowest dynamically stable metastable states are shown in Fig.~$\ref{fig3}$. They gradually
disappear as the parameter $\gamma^2$ approaches $0$, in the sense that local minima in
Fig.~\ref{fig1} cease to exist at this point. This means that metastable phases literally emerge in
Bose-Einstein condensates with 3D SOC only. The presence of metastable phases along with the
true ground states creates the opportunity to simulate false-vacuum decay. Proposed by
Sydney Coleman for modeling phase transitions in the universe \cite{COL77}, decay from a false vacuum into a true one plays a key role in numerous
physical contexts. For example, it occurs in a superheated liquid, where the false vacuum is the liquid
state, while the true one is gaseous \cite{Schmelzer2006}. Thermodynamic fluctuations trigger the continuous appearance of
vapor bubbles in the liquid. Eventually growing bubbles swallow the entire  system. More speculative
manifestations of the phenomena exist also in modern cosmology~\cite{Guth1984,Bousso2013}. Due to the its
high tunability, our system provides an easy route toward testing the false-vacuum quantum decay. The
system can be prepared initially in one of the metastable phases of Fig.~\ref{fig3}. Quantum fluctuations
are then expected to trigger quantum decay accompanied with nucleation of bubbles of one of the
lower-lying true ground states.

\textit{Elementary excitations.\/} ---
The phases and phase transitions in our system can be probed by studying the spectrum of elementary
excitations, e.g., by using Bragg spectroscopy~\cite{MOT12,CHI14,PAN14,KHA14}. Here we consider
the elementary excitations around the PW-Axial ground state; partly motivated by the fact that, for the
case of one-dimensional SOC, interesting roton-like modes were found~\cite{YUN12}. Physically, the
roton mode signals a system's tendency to undergo a first-order phase transition to a  supersolid when
the roton gap closes~\cite{YUN12,YUN13}, and it is usually the consequence of strong correlations in the
system due to the interplay of  SOC and interactions. Our aim is to show that these features persist also
in the case of 3D SOC and that it probes the rich phase diagram obtained above.

The PW-Axial phase has one condensation momentum lying in the $xy$ plane. Without loss of generality,
we choose the condensate momentum to be $\bm{\mathrm{\kappa}}=\frac{\lambda}{2}(-1,0,0)$. Within the
framework of imaginary-time functional integration, the partition function of the system reads~\cite{SIM06}
$\mathcal{Z}=\int \mathcal{D}[\Psi^*,\Psi]\exp{(-S[\Psi^*,\Psi])}$ with the action $S[\Psi^*,\Psi]=\int_0^\beta
d\tau[\int d^3 r \, \sum_\sigma\Psi_\sigma^*\partial_\tau \Psi_\sigma+H-\mu N]$, where $\beta=1/T$
is the inverse temperature, and $\mu$ is the chemical potential introduced to fix the total particle number.
The Bose field is split into the mean-field and fluctuating parts, $\Psi_{\mathbf{q}\sigma}=\left.\Phi_{0\sigma}
\right|_{\mathbf{q}=\bm{\mathrm{\kappa}}}+\phi_{\mathbf{q}\sigma}$. We then expand the action of the
system up to the quadratic order in fluctuating fields obtaining an effective action $S_{\rm eff}\simeq
S_{0}+S_g$. Here $S_{0}=\mathcal{V}\sum_\sigma\left[(-\frac{\lambda^2}{4}-\mu)n_{0\sigma}+
(g+g_{\uparrow\downarrow})n_{0\sigma}^2\right]$ is the mean-field contribution, while $S_g=\frac{1}{2}
\Phi_\mathbf{q}^\dagger\mathcal{G}^{-1}\Phi_\mathbf{q}$ is the fluctuating contribution with a vector field
$\Phi_\mathbf{q}=(\phi_{\vec{\kappa}+\mathbf{q}\uparrow},\phi_{\vec{\kappa}+\mathbf{q}\downarrow},
\phi_{\vec{\kappa}-\mathbf{q}\uparrow}^*, \phi_{\vec{\kappa}-\mathbf{q}\downarrow}^*)^T$.
$\mathcal{G}^{-1}$ is the inverse Green's function of the elementary excitations defined as
\begin{equation}
\mathcal{G}^{-1}=
\begin{pmatrix}
    -iw_n+\epsilon_\mathbf{q}^+ &   R_{\mathbf{q}} &  gn_0 & g_{\uparrow\downarrow}n_0\\
    R_\mathbf{q}^* & -iw_n+\epsilon_\mathbf{q}^- & g_{\uparrow\downarrow}n_0 & gn_0\\
    gn_0 & g_{\uparrow\downarrow}n_0  &  iw_n+\epsilon_{-\mathbf{q}}^+  & R_{-\mathbf{q}}^*\\
    g_{\uparrow\downarrow}n_0   & gn_0  & R_{-\mathbf{q}}&  iw_n+\epsilon_{-\mathbf{q}}^-
\end{pmatrix},
\end{equation}
where $\epsilon_{\mathbf{q}}^\pm=q^2+\frac{\lambda^2}{2}+\lambda(\pm \gamma q_z-q_x)+gn_0$ and $R_{\mathbf{q}}=g_{\uparrow\downarrow}n_0+\lambda(-\frac{\lambda}{2}+q_x-iq_y)$.

\begin{figure}[b]
\includegraphics[width=\columnwidth]{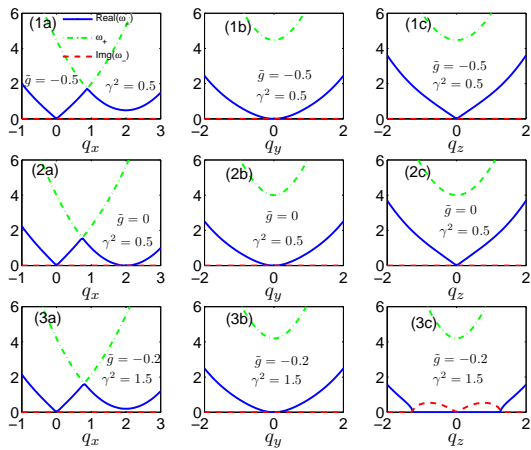}
\caption{(color online) Dispersion of low-lying elementary excitations $\omega_\pm$ along $q_x$
(left panels), $q_y$ (centre panels), and $q_z$ (right panels). First row: $\tilde{g}=-0.5$ and
$\gamma^2=0.5$ (SP-Polar phase is stable), a phonon-maxon-roton feature is seen along the $x$
direction. Second row: $\tilde{g}=0$ and  $\gamma^2=0.5$ (boundary between PW-Axial and
SP-Axial phases), the roton minimum goes soft. Third row: $\tilde{g}=-0.2$ and $\gamma^2=1.5$
(metastability of the PW-Axial phase is broken), imaginary parts appear in $q_z$ dispersion.
We have set $gn_0/\lambda^2=0.25$ in all panels.}
\label{fig4}
\end{figure}

The spectrum of the elementary excitations is determined from the poles of the Green's function.
There are two branches of excitations found as illustrated in Fig.~\ref{fig4}. We choose parameters
to probe the PW-Axial ground state and show the results in the first row in Fig.~\ref{fig4}. The lower
branch of the spectrum exhibits a typical linear Bogoliubov slope at low momenta, followed by a
roton and maxon features at higher momenta.
This structure of the spectrum persists whenever the
momentum ${\mathbf q}$ has components along the axis of the condensation, in this case
the $x$-axis. The roton-maxon feature is absent along any direction that is perpendicular to the
direction of condensation momentum. There is a conical intersection at around $q_x=1$, reflecting the
time-reversal symmetry of the system. It can be lifted by a Zeeman-like
field in the Hamiltonian, in which case the lower branch will become separated from the upper one.
Our purpose here is to study how the spectrum changes when we drive the system across the phase
diagram.

When the system is driven close to the boundary with the SP-Axial phase, the roton minimum becomes
soft as shown in the second row of Fig.~$\ref{fig4}$. This signals the instability of the system toward
the striped order, whose spatial modulation is set by the momentum at which the gap closes. There are
no metastable phases at this boundary, as shown in Fig.~\ref{fig3}. This phase transition is of second
order. On the other hand, nothing dramatic occurs when we move close to the boundary with the SP-Polar
phase. The roton gap does not close, and the spectrum of the PW-Axial phase does not show any specific feature at the phase boundary.
This is due to the  presence of metastable phases. The PW-axial phase becomes metastable when
we cross the line $\gamma^2=1$ at fixed $\tilde{g}<0$ from below, as shown in Fig.~\ref{fig3}. Therefore,
this phase transition is of the first order. We need to drive the system much further to see changes in the
excitation spectrum, namely until the point when metastability breaks down as it is shown in the third row
of Fig.~\ref{fig4}.
The spectrum of the SP phase is qualitatively different from the PW-axial phase and features a double-gapless band structure due to spontaneously broken translations symmetry as shown in Ref.~\cite{YUN13}.
Therefore, measuring the excitation spectrum can be used as a probe of the rich phase
diagram in the presence of 3D SOC.

It is interesting to note that at the tetracritical point $Q_c$, we find also two gapless Goldstone modes,
resulting however from spontaneous breaking of spin-rotation symmetry and U(1) gauge symmetry.
Such modes are expected to remove the four-fold degeneracy at $Q_c$ found at the mean-field level,
leading to a unique ground state via the so called ``order from disorder" mechanism~\cite{HU12,CON11}.

\textit{Experimental relevance.\/} ---
For a  trapped Bose gas in the presence of Weyl SOC and weak inter-particle interaction, one expects
that the ground state is a Skyrmion, which is a superposition of few lowest Landau levels~\cite{KAW12}.
Our predictions should apply for flat bottom traps as in Ref.~\cite{Gaunt2013}. In addition, we may expect the main features of the presented phase diagram be present in harmonic traps with sufficiently strong nonlinearity
\cite{CON13}.


R.~L. acknowledges funding from the NSFC under Grants No. 11274064 and NCET-13-0734.
O.~F. was supported by the Marsden Fund (contract MAU1205), administered by the Royal Society of
New Zealand.

%

\end{document}